\pdfoutput=1
 \documentclass[aps,preprint,nofootinbib]{revtex4}%
\usepackage{amsfonts}
\usepackage{amsmath}
\usepackage{amssymb}
\usepackage{graphicx}%
\providecommand{\U}[1]{\protect\rule{.1in}{.1in}}

\ifx\pdfoutput\relax\let\pdfoutput=\undefined\fi
\newcount\msipdfoutput
\ifx\pdfoutput\undefined\else
\ifcase\pdfoutput\else
\ifx\paperwidth\undefined\else
\ifdim\paperheight=0pt\relax\else\pdfpageheight\paperheight\fi
\ifdim\paperwidth=0pt\relax\else\pdfpagewidth\paperwidth\fi
\fi\fi\fi

\begin{document}
 \preprint{USC-HEP-10B1}
\title{Gauge Symmetry in Phase Space \\Consequences for Physics and Spacetime }
\thanks{Work partially supported by the US Department of Energy, grant number DE-FG03-84ER40168.}
\author{Itzhak Bars}
\affiliation{Department of Physics and Astronomy, University of
Southern California, Los Angeles, CA 90089-0484, USA}
\author{{\textbf{Gell-Mann: Anything which is not forbidden is
compulsory!
\\ This paper is dedicated to Murray Gell-Mann for his 80th
birthday.}} }
\thanks{Lecture at the Conference in Honor of Murray Gell-Mann's 80th Birthday, to
appear in IJMPA.}

\begin{abstract}
{\footnotesize Position and momentum enter at the same level of
importance in the formulation of classical or quantum mechanics.
This is reflected in the invariance of Poisson brackets or quantum
commutators under canonical transformations, which I regard as a
global symmetry. A gauge symmetry can be defined in phase space
}$(X^{M},P_{M})${\footnotesize that imposes equivalence of momentum
and position for every motion at every instant of the worldline. One
of the consequences of this gauge symmetry is a new formulation of
physics in spacetime. Instead of one time there must be two, while
phenomena described by one-time physics in 3+1 dimensions appear as
various \textquotedblleft shadows\textquotedblright\ of the same
phenomena that occur in 4+2 dimensions with one extra space and one
extra time dimensions (more generally, d+2). The 2T-physics
formulation leads to a unification of 1T-physics systems not
suspected before and there are new correct predictions from
2T-physics that 1T-physics is unable to make on its own
systematically. Additional data related to the predictions, that
provides information about the properties of the extra 1-space and
extra 1-time dimensions, can be gathered by observers stuck in 3+1
dimensions. This is the probe for investigating indirectly the extra
1+1 dimensions which are neither small nor hidden. This 2T formalism
that originated in 1998 has been extended in recent years from the
worldline to field theory in d+2 dimensions. This includes 2T field
theories that yield 1T field theories for the Standard Model and
General Relativity as shadows of their counterparts in 4+2
dimensions. Problems of ghosts and causality in a 2T space-time are
resolved automatically by the gauge symmetry, while a higher
unification of 1T field theories is obtained. In this lecture the
approach will be described at an elementary worldline level, and the
current status of 2T-physics will be summarized.}
\end{abstract}
\bigskip

 \maketitle

\section{Some consequences of the gauge symmetry}

Gauge symmetry in phase space is an unfamiliar concept. What
motivates it? In my own work the motivation emerged from my
observation in 1995 that the 11-dimensional extended supersymmetry
(SUSY) in M-theory is really a 12-dimensional SUSY with an SO$\left(
10,2\right)  $ symmetry, indicating the possibility of a 12D
spacetime with two-time (2T) signature \cite{Stheory} (see also
F-theory in 12D that developed soon afterwards \cite{ftheory}, and
S-theory in 11+2 dimensions \cite{Stheory}). However, if taken
seriously, a theory in such a spacetime would be riddled with
problems of ghosts and causality. If such a spacetime is more than a
mathematical accident in M- or F- or S- theory, one would have to
understand how to construct a physical theory that overcomes these
problems. In 1995 the challenge seemed to be worthwhile in its own
right, in addition to possibly providing a guide for constructing M-
or F- or S- theory and explaining the origin of the SO$\left(
10,2\right)  $ or SO$\left(  11,2\right)  $ symmetry.

Yesterday, Murray Gell-Mann talked about ``Some Lessons from 60
Years of Theorizing". One of his main messages was that from time to
time we should question some of the ``received ideas". Usually, he
said, there are good reasons for why they exist, but they are
sometimes wrong. He gave examples in many areas of ``received ideas"
which turned out to be wrong, including some famous ones such as the
Earth being the center of the universe. Then he elaborated on how he
overcame some of those received ideas in his own research, leading
to some of his major successes that we are celebrating in this
conference. Such refreshing words from our energetic honoree lifted
my spirits and reaffirmed my admiration for his intellect and his
seminal work.

Well, that spacetime has only one time coordinate is one of those
received ideas. A major argument in favor of this one is that
apparently insurmountable problems with ghosts\ and causality
prevent additional timelike dimensions. I questioned this received
idea in 1995 and three years later, in 1998, found how to overcome
it. The key is the gauge symmetry in phase space
\cite{2treviews}$^{,}$\cite{2tbacgrounds} that I will discuss in
this talk.

The resolution of similar problems in one-time theories taught us over the
past century that the solution to the ghost problem associated with the first
time dimension is to have some carefully constructed gauge symmetries. Gauge
symmetries in general relativity, Maxwell or Yang-Mills theories, as well as
string theory are essential to remove ghosts, thus providing a physically
sensible theory. A gauge symmetry has a dual role. On the one hand it is the
very reason for the existence of the fundamental forces while dictating the
form of fundamental equations of physics, on the other hand it removes ghosts.
It was evident to me that, to remove the ghost and causality problems, that
are the stumbling blocks in a theory with two times, a much stronger gauge
symmetry was needed. Furthermore, if such a thing existed it would lead to
some powerful constraints on the fundamental formulation of physics. This
could also be a guiding principle for constructing M-theory.

Before I describe the phase space gauge symmetry based on symplectic
transformations Sp$\left(  2,R\right)  $ let me highlight some of its
important consequences.

\begin{itemize}
\item The Sp$\left(  2,R\right)  $ gauge symmetry requires that all physics be
reformulated in 4+2 dimensions (more generally d+2). 2T is a consequence, not
an input. Thus, for phase space $\left(  X^{M},P_{M}\right)  ,$ and all fields
$A_{M}\left(  X\right)  ,G_{MN}\left(  X\right)  ,$ etc., 2T signature is
required by the symmetry, not just permitted. The underlying Sp$\left(
2,R\right)  $ leads to greater gauge symmetry and constraints that remove all
ghosts or causality problems. I called this 2T-physics.

\item All 1T physics for which we have experimental evidence so far, at all
known scales of energy or distance, fits into 2T-physics. The gauge
invariant sector of 2T-physics in 4+2 dimensions, namely the ghost
free physical sector, becomes effectively a one-time (1T) theory
with an effective 3+1 dimensions. There remains no Kaluza-Klein type
degrees of freedom at all. But the outcome is not the same as the 1T
formulation of physics. Finding again 3+1 within 4+2 is not a zero
sum game, because there are many ways in which 3+1 \textit{phase
space} is embedded in 4+2 \textit{phase space}, leading to many
emergent ``times" and corresponding ``Hamiltonians" within 4+2. I
call the emergent 3+1
spacetimes and dynamical systems \textquotedblleft shadows\textquotedblright%
\ of the \textquotedblleft substance\textquotedblright\ in 4+2. This leads
systematically to a large number of correct predictions by 2T-physics, in the
form of hidden relations between dynamical systems and hidden symmetries in
3+1 dimensions, that the standard 1T formulation of physics (1T-physics) is
not equipped to predict but can only verify. The new information in 3+1
provided by the systematic predictions from 4+2 (more generally d+2) is the
main new content of 2T-physics.

\item 2T-physics was initially formulated as a theory for particles moving on
worldlines. In recent years the formulation was successfully
extended to 2T field theory, which includes the Standard Model (SM)
\cite{2tstandardM} and General Relativity (GR)
\cite{2tGravity}$^{,}$\cite{2tGravDetails} as 2T field theories in
4+2 dimensions. These are consistent with their 1T counterparts in
3+1 dimensions. In fact the usual SM and GR emerge as one of the
shadows from 4+2, namely the \textquotedblleft conformal
shadow\textquotedblright. The status of further developments of the
2T approach, including SUSY, higher dimensions, string theory, will
be summarized at the end. Suffice it to say that 2T-physics agrees
with 1T-physics but it goes beyond by its potential to make new
testable predictions, that 1T-physics misses and, which so far are
consistent with known data. This additional information, namely the
hidden symmetries and the systematic relationships among the
emergent multiple shadows, provides a probe for discovering
indirectly the properties of the extra 1+1 dimensions.
\end{itemize}

\section{Phase space gauge symmetry Sp$\left(  2,R\right)  $}

A clue for the fundamental principle is a \textit{position}$\leftrightarrow
$\textit{momentum} global symmetry in classical or quantum mechanics.
Specifically, position and momentum appear at the same level of importance in
specifying boundary conditions or in reporting the results of any measurement.
More importantly the formulation of classical mechanics in terms of Poisson
brackets $\left\{  X^{M},P_{N}\right\}  =\delta_{N}^{M}$ is invariant under
all infinitesimal canonical transformations, $\delta_{\varepsilon}X^{M}%
=\frac{\partial\varepsilon\left(  X,P\right)  }{\partial P_{M}},$
$\delta_{\varepsilon}P_{M}=-\frac{\partial\varepsilon\left(  X,P\right)
}{\partial X^{M}}$, since $\delta_{\varepsilon}\{X^{M},P_{N}\}=0$ for any
$\varepsilon\left(  X,P\right)  $. A quantum ordered version of the same
symmetry holds for the fundamental quantum commutators $[X^{M},P_{N}%
]=i\hbar\delta_{N}^{M}$, since $\delta_{\varepsilon}[X^{M},P_{N}]=0.$ The
symmetry under infinitesimal classical canonical transformations is also the
symmetry of the \textit{first term} of any action in the first order formalism
$S=\int d\tau(\dot{X}^{M}P_{M}-\cdots)$ since one gets a total derivative for
\begin{equation}
\delta_{\varepsilon}\left(  \dot{X}^{M}P_{M}\right)  =\frac{d}{d\tau}\left(
P_{M}\frac{\partial\varepsilon\left(  X,P\right)  }{\partial X^{M}%
}-\varepsilon\left(  X,P\right)  \right)  .
\end{equation}

This symmetry is spoiled for the action when a specific Hamiltonian $H\left(
X,P\right)  $ is inserted in the action as part of the $``\cdots
$\textquotedblright. However a specific Hamiltonian focusses on a specific
dynamical system rather than the general formalism. We learned in special and
general relativity that the notion of time and the corresponding Hamiltonian
are dependent on the observer. Einstein showed us how to detach the
formulation of fundamental laws from the perspective of observers by requiring
equivalence of all perspectives in all spacetime frames. My idea was to take
this equivalence notion one step further to all perspectives in phase space
$\left(  X^{M},P_{M}\right)  $, not only perspectives in spacetime $X^{M}$, by
requiring a gauge symmetry in phase space.

In this approach the Hamiltonian (and the associated time) would be regarded
as an emergent concept that depends on some perspective from the point of view
of phase space. Hence, I ignored the Hamiltonian and instead focused on
requiring an action principle with a local symmetry in phase space.

To begin, the canonical transformation above should be regarded as a
\textit{global} symmetry on the worldline since the $\varepsilon\left(
X,P\right)  $ of canonical transformations depends on the proper time $\tau$
only through the \textquotedblleft fields\textquotedblright\ $X\left(
\tau\right)  ,P\left(  \tau\right)  $. So any infinitesimal parameters
included in $\varepsilon\left(  X,P\right)  $ are global parameters. To have a
local symmetry on the worldline one needs a symmetry with parameters that
depend arbitrarily on the worldline parameter $\tau$, through additional
$\tau$ dependent parameters, leading to $\varepsilon\left(  X\left(
\tau\right)  ,P\left(  \tau\right)  ,\tau\right)  $ local on the worldline$.$
It turns out that there is a limit on how large the symmetry can be because
the system may turn out to be trivial if constrained by too much local
symmetry. What worked is an Sp$\left(  2,R\right)  $ local
symmetry\footnote{This is for a spinless particle. For particles with spin
and/or supersymmetry the symmetry group is larger, but it must include
Sp$\left(  2,R\right)  $ as a subgroup in a special way \cite{spin2t}%
\cite{twistorLect}\cite{2treviews}.\label{withspin}} formulated as
follows.

Introduce the three generators of Sp$\left(  2,R\right)  $ as a symmetric
2$\times2$ tensor $Q_{ij}\left(  X,P\right)  ,$ namely $Q_{11}\left(
X,P\right)  ,Q_{22}\left(  X,P\right)  $ and $Q_{12}\left(  X,P\right)
\equiv$ $Q_{21}\left(  X,P\right)  $ and require that they form the Lie
algebra of Sp$\left(  2,R\right)  $ under Poisson brackets. I will then
require local symmetry on the worldline with arbitrary local prameters
$\omega^{ij}\left(  \tau\right)  $ that define $\varepsilon\left(
X,P,\tau\right)  =\frac{1}{2}\omega^{ij}\left(  \tau\right)  Q_{ij}\left(
X,P\right)  .$ An action invariant under this local transformation can be
constructed by introducing the Sp$\left(  2,R\right)  $ gauge potentials on
the worldline $A^{ij}\left(  \tau\right)  $
\begin{equation}
S=\int d\tau\left(  \dot{X}^{M}P_{M}-\frac{1}{2}A^{ij}\left(  \tau\right)
Q_{ij}\left(  X\left(  \tau\right)  ,P\left(  \tau\right)  \right)  \right)  .
\label{action}%
\end{equation}
It can be verified that the action is invariant under the local
transformations of the matter and gauge degrees of freedom $\delta
_{\varepsilon}X^{M}=\frac{1}{2}\omega^{ij}\left(  \tau\right)  \frac{\partial
Q_{ij}\left(  X,P\right)  }{\partial P_{M}},$ $\delta_{\varepsilon}%
P_{M}=-\frac{1}{2}\omega^{ij}\left(  \tau\right)  \frac{\partial Q_{ij}\left(
X,P\right)  }{\partial X^{M}}$, $\delta_{\varepsilon}A^{ij}=D_{\tau}%
\omega^{ij}\equiv\partial_{\tau}\omega^{ij}+\left[  A,\omega\right]  ^{ij}$
(summed indices are contracted with the antisymmetric Sp$\left(  2,R\right)  $
metric $\varepsilon_{ij}$), provided the $Q_{ij}$ form the Lie algebra under
Poisson brackets. It is possible to generalize this action by adding a term of
the form $S^{\prime}=-\int d\tau U\left(  X\left(  \tau\right)  ,P\left(
\tau\right)  \right)  $ provided $U\left(  X,P\right)  $ is invariant under
Sp$\left(  2,R\right)  ,$ namely $\left\{  Q_{ij},U\right\}  =0.$

The equivalence principle in phase space I outlined suggests that one should
consider all possible $Q_{ij}\left(  X,P\right)  $ that satisfy Sp$\left(
2,R\right)  $ to recover all possible physical systems for a spinless particle
(for particles with spin see footnote \ref{withspin}). I found an infinite
number of $Q_{ij}\left(  X,P\right)  $ that form Sp$\left(  2,R\right)  $ and
I classified them up to canonical transformations \cite{2tbacgrounds}%
\cite{noncommutative}. I now consider some examples.

An example of the $Q_{ij}\left(  X,P\right)  $ that satisfy Sp$\left(
2,R\right)  $ is%
\begin{equation}
\text{Example:\ }Q_{11}=X\cdot X,\;Q_{22}=P\cdot P,\;Q_{12}=X\cdot P.
\label{special}%
\end{equation}
These special $Q_{ij}$ are constructed by using a dot product
$X\cdot X=X^{M}X^{N}\eta_{MN}$ where the signature of the flat
metric $\eta_{MN}$ in target space is not specified \textit{\`{a}
priori}. The Sp$\left( 2,R\right)  $ invariants that satisfy
$\left\{ Q_{ij},U\right\}  =0$ are all possible functions $U\left(
L^{MN}\right)  $ of the angular momentum generators
$L^{MN}=X^{M}P^{N}-X^{N}P^{M}.$ For this example the Sp$\left(
2,R\right)  $ transformation defined above through Poisson brackets
amounts to a \textit{local} linear transformation on $\left(
X^{M},P^{M}\right)  $ such
that these behave like a doublet for each $M,$ as follows \cite{2treviews}%
\begin{equation}
\left(
\begin{array}
[c]{c}%
X^{\prime M}\left(  \tau\right) \\
P^{\prime M}\left(  \tau\right)
\end{array}
\right)  =\left(
\begin{array}
[c]{cc}%
a\left(  \tau\right)  & b\left(  \tau\right) \\
c\left(  \tau\right)  & d\left(  \tau\right)
\end{array}
\right)  \left(
\begin{array}
[c]{c}%
X^{M}\left(  \tau\right) \\
P^{M}\left(  \tau\right)
\end{array}
\right)  ,\;ad-bc=1,
\end{equation}
where for the infinitesimal transformation $a\left(  \tau\right)
=1+\omega^{12}\left(  \tau\right)  +\cdots$,$\;b\left(  \tau\right)
=\omega^{22}\left(  \tau\right)  +\cdots$,$\;c\left(  \tau\right)
=-\omega^{11}\left(  \tau\right)  +\cdots$,$~d\left(  \tau\right)
=1-\omega^{12}\left(  \tau\right)  +\cdots\ .\;$Furthermore, the action above
can be rewritten in terms of usual Yang-Mills type covariant derivatives
appropriate for the doublets $\left(  X,P\right)  $\cite{2treviews}.

One of the consequences of the general action (\ref{action}) is the equation
of motion for the gauge field $A^{ij}$ which acts like a Lagrange multiplier.
This requires that the Sp$\left(  2,R\right)  $ charges should vanish
\begin{equation}
Q_{ij}\left(  X,P\right)  =0.
\end{equation}
The meaning of this equation is that only the Sp$\left(  2,R\right)  $ gauge
invariant subspace of phase space is physical. Hence, only gauge invariant
motion is allowed. The solution space for these Sp$\left(  2,R\right)  $
conditions are called \textquotedblleft shadows\textquotedblright. I will show
some examples of shadows in the next section.

It turns out that nontrivial solutions to $Q_{ij}=0$ exist only if the target
space has 2 times, no less and no more. To see why, consider the example in
Eq.(\ref{special}). If the metric $\eta_{MN}$ is Euclidean (0T) then the only
solution is trivial $X^{M}=P^{M}=0.$ If the metric $\eta_{MN}$ is Minkowski
(1T) then a solution is possible only if $X^{M},P^{M}$ are lightlike and
parallel, which means the angular momentum vanishes $L^{MN}=0.$ This is
trivial because it does not describe even a free particle. To have non-trivial
solutions one must have a metric $\eta_{MN}$ with two times (2T) or more. With
2T it turns out there is just enough gauge symmetry to remove the ghosts, but
with three or more times the Sp$\left(  2,R\right)  $ gauge symmetry is
insufficient to remove ghosts. Hence there must be two times, no less and no
more. I have shown that 2T is an outcome, not an input, since it is demanded
by the gauge symmetry and nontrivial physical content.

For the model of 2T-physics based on the $Q_{ij}$ in Eq.(\ref{special}) there
is an automatic \textit{global} SO$\left(  d,2\right)  $ symmetry. This
SO$\left(  d,2\right)  $ is the symmetry of the dot products and has
generators $L^{MN}$ that are Sp$\left(  2,R\right)  $ gauge invariant
$\left\{  Q_{ij},L^{MN}\right\}  =0,$ with $L^{MN}=X^{M}P^{N}-X^{N}P^{M}.$ The
action may be modified by an additional Sp$\left(  2,R\right)  $ gauge
invariant term of the form $S^{\prime}=-\int d\tau U\left(  L\left(
\tau\right)  \right)  ,$ where $U\left(  L\right)  $ is an arbitrary function
of the $L^{MN}$, which could break the global SO$\left(  d,2\right)  $
symmetry partially or fully. The inclusion of $U$ does not change the
essential point that the $Q_{ij}$ must vanish, leading to the same 1T shadows
(see next section), and that spacetime is $d+2$ at the fundamental level, even
if the global SO$\left(  d,2\right)  $ symmetry is broken.

I am often asked if it is possible to have more times by enlarging
the gauge symmetry beyond Sp$\left(  2,R\right)  .$ My answer is
that it is unlikely, but I don't have a theorem so far. This is
based on the following considerations. First, it is certainly
possible to write a gauge invariant action identical in form to
(\ref{action}) for any Lie algebra whose generators $Q_{a}\left(
X,P\right)  $ close (assuming these can be constructed in phase
space). The issue is whether the gauge invariance condition
$Q_{a}\left(  X,P\right)  =0$ has non-trivial content and also if
the emerging shadows are ghost free. In all attempts so far, with
specific examples $Q_{a}\left(  X,P\right)  $ for spinless
particles, we have found that, such a scheme based on noncompact
groups, either leads to trivial content for the solutions of
$Q_{a}=0,$ or the emergent spaces (i.e. shadows) have ghosts because
all the timelike dimensions could not be removed from $X^{M},P_{M}$.
One remarkable exception that has worked so well is Sp$\left(
2,R\right)  ,$ which seems to indicate that 2T-physics may be
special \cite{no3t}.

The consequences of the local worldline Sp$\left(  2,R\right)  $ symmetry for
local field theory (fields that depend only on $X^{M}$) \cite{2tfield}%
\cite{emergentfieldth2} and an extension of these concepts to field
theory in phase space (fields that depend on both $X$ and $P$)
\cite{noncommutative} have been developed, but there will not be
sufficient time to discuss them in this talk. They will be described
only briefly at the end of this paper.

\section{Shadows}

In this section I concentrate on the \textit{2T free particle in flat
spacetime} described by the $Q_{ij}$ in Eq.(\ref{special}). To obtain the 1T
shadows I will make two gauge choices and solve the two constraints $X^{2}=0$
and $X\cdot P=0.$ This fixes two components of $X^{M}$ and two components of
$P^{M}$ in terms of the remaining independent degrees of freedom, thus
reducing the theory from $d+2$ dimensions to various shadows in $d$
dimensions. There will remain still one gauge symmetry and one unsolved
constraint that can remove the ghosts in the remaining timelike degree of
freedom in the shadow.

To perform these steps it is useful to define a lightcone type basis
$X^{M}=(X^{+^{\prime}},X^{-^{\prime}},X^{\mu})$ so that the flat metric in
$d+2$ dimensions is expressed as $ds^{2}=-2dX^{+^{\prime}}dX^{-^{\prime}%
}+dX^{\mu}dX^{\mu}\eta_{\mu\nu}$ with $\eta_{\mu\nu}$ the Minkowski metric in
$d$ dimensions including 1 time.

As the first example of a shadow, consider the following solution
\cite{2treviews} which I call the \textquotedblleft\textit{conformal
shadow}\textquotedblright\
\begin{equation}%
\begin{array}
[c]{lll}%
X^{+^{\prime}}\left(  \tau\right)  =1,\; & X^{-^{\prime}}=\frac{1}{2}%
x^{2}\left(  \tau\right)  ,\; & X^{\mu}\left(  \tau\right)  \equiv x^{\mu
}\left(  \tau\right) \\
P^{+^{\prime}}\left(  \tau\right)  =0, & P^{-^{\prime}}=x\left(  \tau\right)
\cdot p\left(  \tau\right)  , & P^{\mu}\left(  \tau\right)  \equiv p^{\mu
}\left(  \tau\right)
\end{array}
,\;p^{2}=0. \label{massless}%
\end{equation}
Here the two gauges are $X^{+^{\prime}}\left(  \tau\right)  =1$ and
$P^{+^{\prime}}\left(  \tau\right)  =0$ for all $\tau.$ The solution of the
constraint $X^{2}=0$ yields $X^{-^{\prime}}$ and the solution of the
constraint $X\cdot P=0$ yields $P^{-^{\prime}}$as given above. The remaining
degrees of freedom which were named as $x^{\mu},p^{\mu}$ are still subject to
the constraint $P^{2}=-2P^{+^{\prime}}P^{-^{\prime}}+P^{\mu}P_{\mu}=0$ which
takes the form $p^{2}=0.$ This phase space $\left(  x^{\mu}\left(
\tau\right)  ,p^{\mu}\left(  \tau\right)  \right)  $ describes the free
massless 1T relativistic particle in $d$ dimensions. This is confirmed by
inserting the gauge choices into the original action, yielding $S=\int
d\tau\left(  \dot{x}^{\mu}p_{\mu}-\frac{1}{2}A^{22}p^{2}\right)  ,$ which is
the action for the free massless relativistic particle. One can also start
from the equations of motion for $X^{M}\left(  \tau\right)  ,P^{M}\left(
\tau\right)  ,$ insert the gauge fixed configuration above, and obtain the
equations of motion of the free massless relativistic particle.

The original action had an SO$\left(  d,2\right)  $ global symmetry. The
global symmetry of the action does not disappear since the action is gauge
invariant. However, it becomes hard to notice the symmetry in terms of the
remaining degrees of freedom because it takes a non-linear form. The
generators $L^{MN}=X^{M}P^{N}-X^{N}P^{M}$ were also gauge invariant, so they
can be expressed in terms of the remaining degrees of freedom by inserting the
configuration in Eq.(\ref{massless}). This gives the following components of
$L^{MN}$ in their shadow form%

\begin{equation}%
\begin{array}
[c]{c}%
L^{+^{\prime}-^{\prime}}=x\cdot p,\;~~L^{\mu\nu}=x^{\mu}p^{\nu}-x^{\nu}p^{\mu
},\\
L^{+^{\prime}\mu}=p^{\mu},\ L^{-^{\prime}\mu}=\frac{1}{2}x^{2}p^{\mu}-x\cdot
px^{\mu}.
\end{array}
\label{masslessLMN}%
\end{equation}
This is recognized as the generators of the conformal group
SO$\left( d,2\right)  $\footnote{Dirac \cite{Dirac} was the first to
use a 6 dimensional space to describe conformal symmetry SO$\left(
4,2\right)  .$ His approach, which was further developed
\cite{salam}\cite{marnelius06} had faded away when I discovered my
approach, \textit{as a gauge symmetry of phase space,} without being
aware of Dirac's different formalism or reasoning. We now know that
Dirac's work is automatically part of 2T-physics, since it coincides
with my \textquotedblleft conformal shadow\textquotedblright\ for
the special case of $Q_{ij}$ in (\ref{special}). But this is just an
example of a particular shadow within the larger scope of
2T-physics.}$.$ Their action on the massless degrees of freedom
$x^{\mu},p^{\mu}$ is given by computing their Poisson brackets
$\delta_{\varepsilon}x^{\mu}=\frac{1}{2}\varepsilon_{MN}\left[
L^{MN},x^{\mu}\right]  $ and similarly for
$\delta_{\varepsilon}p^{\mu}$. Using this, one can check that these
are indeed generators of symmetry for the action for the massless
particle \cite{2treviews}\cite{2tHandAdS}. This is expected
automatically since both $S$ and $L^{MN}$ are gauge invariants and
one already knew that $S$ was invariant under SO$\left(  d,2\right)
.$

A second example is the massive relativistic particle given by the
following shadow configuration (a different looking form in
\cite{2tHandAdS} is gauge
equivalent)%
\begin{equation}%
\begin{array}
[c]{ccc}%
X^{+^{\prime}}=\frac{1+a}{2a},\; & X^{-^{\prime}}=\frac{x^{2}a}{1+a},\; &
X^{\mu}\equiv x^{\mu}\left(  \tau\right) \\
P^{+^{\prime}}=\frac{-m^{2}}{2ax\cdot p}, & P^{-^{\prime}}=ax\cdot p, &
P^{\mu}\equiv p^{\mu}\left(  \tau\right)
\end{array}
,\;%
\genfrac{}{}{0pt}{}{a\equiv\left(  1+\frac{m^{2}x^{2}}{\left(  x\cdot
p\right)  ^{2}}\right)  ^{1/2}}{p^{2}+m^{2}=0}
\label{massiveRel}%
\end{equation}
The remaining constraint in this case $P^{2}=0$ takes the form $p^{2}%
+m^{2}=0,$ which says that the shadow phase space $\left(
x^{\mu}\left( \tau\right)  ,p^{\mu}\left(  \tau\right)  \right)  $
now corresponds to the 1T massive relativistic particle. As before,
this can be confirmed by computing both the action and the equations
of motion. Now we get a surprise not noticed before in 1T-physics.
The 2T approach leads us to expect that the action for the massive
relativistic particle $S=\int d\tau\left(  \dot{x}^{\mu}p_{\mu
}-\frac{1}{2}A^{22}\left(  p^{2}+m^{2}\right)  \right)  ,$ which is
a gauge fixed form of the original action (\ref{action}), should be
invariant under SO$\left(  d,2\right)  ,$ but no one suggested this
before in 1T-physics. To find out how to construct the symmetry
generators in this case (the massive analog of (\ref{masslessLMN}))
\cite{2tHandAdS} all one needs to do is insert
the shadow phase space of Eq.(\ref{massiveRel}) into $L^{MN}=X^{M}P^{N}%
-X^{N}P^{M}$.

Even more surprising (for 1T-physics) in this regard is the shadow for the
massive non-relativistic particle given by \cite{2tHandAdS}
\begin{equation}%
\begin{array}
[c]{llll}%
X^{+^{\prime}}=t\left(  \tau\right)  ,\; & X^{-^{\prime}}=\frac{\vec{r}%
\cdot\vec{p}-tH}{m},\; & X^{0}=\pm\left\vert \vec{r}-\frac{t}{m}\vec
{p}\right\vert ,\;\; & X^{i}=\vec{r}^{i}\left(  \tau\right) \\
P^{+^{\prime}}=m, & P^{-^{\prime}}=H\left(  \tau\right)  , & P^{0}=0 &
P^{i}=\vec{p}^{i}\left(  \tau\right)
\end{array}
\label{nonrelativistic}%
\end{equation}
where $t\left(  \tau\right)  $ and $H\left(  \tau\right)  $ are shadow
canonical variables just like $\left(  \vec{r}\left(  \tau\right)  ,\vec
{p}\left(  \tau\right)  \right)  $ as described by the action $S=\int
d\tau\left(  -H\partial_{\tau}t+\partial_{\tau}\vec{r}\cdot\vec{p}-\frac{1}%
{2}A^{22}\left(  -2mH+\vec{p}^{2}\right)  \right)  ,$ which follows from
(\ref{action}) by inserting the shadow configuration above. In this case the
remaining constraint $P^{2}=0$ takes the form $-2mH+\vec{p}^{2}=0$ which shows
that $H$ is the non-relativistic Hamiltonian when the constraint is solved and
the final gauge choice is made $t\left(  \tau\right)  =\tau.$ The
corresponding completely gauge fixed action in $\left(  d-1\right)  $
\textit{space} dimensions is $S=\int d\tau\left(  \partial_{\tau}\vec{r}%
\cdot\vec{p}-\frac{\vec{p}^{2}}{2m}\right)  .$ Evidently it
describes the massive nonrelativistic particle. However,
surprisingly (for 1T-physics) it is invariant under SO$\left(
d,2\right)  ,$ which is realized by rather complicated non-linear
and $\tau$ dependent transformations generated by $L^{MN},$ which
are the analogs \cite{2tHandAdS} of Eq.(\ref{masslessLMN}).

Another remarkable property is the emergence of the mass parameter $m$ in
Eqs.(\ref{massiveRel}) and (\ref{nonrelativistic}) as a modulus in the
embedding of the shadow phase space $\left(  x,p\right)  $ in the higher
dimensional phase space $\left(  X,P\right)  $. I still believe that mass is
very likely explained by the Higgs particle as hopefully will be confirmed at
the LHC. But the alternative mass generation mechanism I have just displayed
must also mean something in Nature and I hope to find its meaning some day. If
the Higgs gets into trouble at the LHC it might be a good idea to investigate
seriously for the origin of mass in this alternative direction.%

\begin{center}
\includegraphics[
natheight=3.802600in,
natwidth=5.061800in,
height=3.8026in,
width=5.0618in
]%
{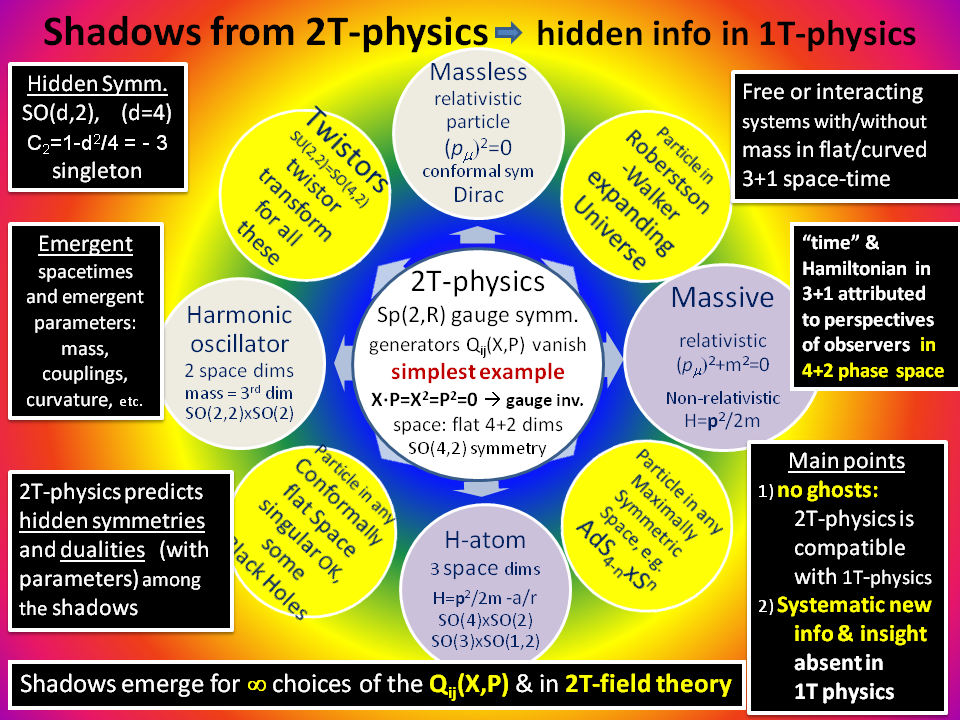}%
\\
Fig.1: Some 1T shadows of the 2T free particle in flat spacetime.
\label{shadowspng}%
\end{center}

So far I displayed shadows of the free 2T particle in flat space
that behave like free particles in 1T-physics. However, there are
also all sorts of shadows of the same \textquotedblleft
substance\textquotedblright\ that behave like particles subject to a
variety of forces, as shown with some examples in Fig.1. These
include some non-relativistic potentials (Hydrogen atom, harmonic
oscillator), and some curved spaces, such as the Robertson-Walker
expanding universe, any conformally flat space
(AdS$^{d-n}\times$S$^{n},$ maximally symmetric space), and even some
singular spaces. Furthermore for all of these shadows new twistor
formulations provide an alternative expression of the shadow phase
space \cite{2ttwistor}\cite{twistorLect}, as indicated on the
figure. The mathematical expressions for these shadows (similar to
Eqs.(\ref{massless},\ref{massiveRel},\ref{nonrelativistic})) were
developed non-systematically \cite{2tHandAdS} over several years and
some of them are summarized in tables I,II,III in
 \cite{emergentfieldth1}.

So, parameters, such as mass, coupling, curvature, emerge from the moduli for
embedding $\left(  x,p\right)  $ into $\left(  X,P\right)  .$ All of these
shadows have the hidden SO$\left(  d,2\right)  $ symmetry, which is realized
in terms of non-linear realizations of $L^{MN}.$ When these systems are
quantized, the Casimir operators can be evaluated (they are all zero at the
classical level) and show that they all give the same quantized value, such as
$C_{2}=\frac{1}{2}L_{MN}L^{MN}=1-d^{2}/4.$ This says that this is the
singleton representation of SO$\left(  d,2\right)  .$ All shadows are in the
same representation, but each shadow is realized in unitarily equivalent bases
of this symmetry.

Evidently there is a lot of information in the hidden relationships among
these systems. This information resides in the gauge invariant properties of
the 2T \textquotedblleft substance\textquotedblright\ in $d+2$ dimensions
which is captured holographically by each 1T shadow in $d$ dimensions.
1T-physics treats all the shadows as different from each other and gives no
clues that they may be related. By contrast 2T-physics makes the prediction
that observers in $d$ dimensions will discover the predicted relationships and
hidden symmetries if they look hard enough.

The relationship between the shadows is similar to duality transformations,
which in the present case amount to Sp$\left(  2,R\right)  $ gauge
transformations from one fixed gauge to another. These transformations involve
not only change of coordinates and momenta but also parameters such as mass,
coupling, curvature, etc. All the relations among shadows amount to the fact
that $L^{MN}$ is gauge invariant and therefore any function $F\left(
L\right)  $ must have the same gauge invariant value in all the shadows as
expressed in terms of the phase space for that shadow. This is the key for all
the expected duality relations derivable from the free 2T particle in flat spacetime.

A complete classification of the possible shadows that emerge from
the set of $Q_{ij}=\left(  X^{2},P^{2},X\cdot P\right)  $ is not
known. Other forms of $Q_{ij}\left(  X,P\right)  $ will produce
their own set of shadows. Similarly the corresponding 2T field
theories \cite{2tfield}\cite{susy2tN2N4}
produce shadows in the form of 1T field theories \cite{emergentfieldth1}%
\cite{emergentfieldth2}. This rich set of dualities is likely to be
useful for developing computational tools. So far this has remained
largely unexplored due to lack of time and other pressing
priorities.

In summary, quite generally, 2T-physics defines a \textquotedblleft
substance\textquotedblright\ that has many \textquotedblleft
shadows\textquotedblright\ in 1T-physics. Each one of them inherits
holographically the gauge invariant properties of the \textquotedblleft
substance\textquotedblright\ (i.e. the theory defined by $Q_{ij}\left(
X,P\right)  $, and corresponding generalization in field theory, including
spin, etc.)\ in $d+2$ dimensions, but the shadows themselves are effective
systems in $\left(  d-1\right)  +1$ dimensions with only 1T. Many possible 1Ts
emerge from phase space in $d+2$ dimensions, so the 1T in a given shadow is
not the same 1T in another shadow. For this reason each shadow is described by
a different Hamiltonian in the usual language of 1T-physics. Automatically,
these 1T dynamical systems are related to each other by their gauge invariance
properties. But 1T-physics is not equipped to display those hidden
relationships among the shadows, because for 1T-physics they seem like
unrelated dynamical systems with separate Hamiltonians. This is how 1T-physics
misses the systematic predictions of 2T-physics.

\section{Gravity and Standard Model in 2T-physics}

A particle moving in arbitrary backgrounds, including
electromagnetic, gravitational or other general fields in $d+2$
dimensions is formulated in terms of more general $Q_{ij}\left(
X,P\right)  $ \cite{2tbacgrounds}. This formulation treats the
Maxwell-type gauge symmetries, general coordinate transformations
and more general cases of gauge symmetries in a unified way, all as
special forms of canonical transformations \cite{2tbacgrounds}. I
consider here just the gravitational background given by \cite{2tbacgrounds}%
$^{,}$\cite{2tGravity}
\begin{equation}
Q_{11}=W\left(  X\right)  ,\;Q_{12}=V^{M}\left(  X\right)  P_{M}%
,\;Q_{22}=G^{MN}\left(  X\right)  P_{M}P_{N}.
\end{equation}
Contrast this to the flat background in Eq.(\ref{special}) to understand the
significance of the background fields, noting that one specializes to the flat
case with $G_{flat}^{MN}\left(  X\right)  =\eta^{MN}$, $W_{flat}\left(
X\right)  =X^{2}$, and $V_{flat}^{M}=X^{M}$. There is one further requirement
for this to be compatible with the Sp$\left(  2,R\right)  $ gauge symmetry of
the action in Eq.(\ref{action}), that is, these $Q_{ij}$ must close into the
Sp$\left(  2,R\right)  $ Lie algebra under Poisson brackets. Consequently the
background fields $W\left(  X\right)  ,V^{M}\left(  X\right)  ,G^{MN}\left(
X\right)  $ must satisfy certain equations which I have called the
\textquotedblleft\textit{kinematic equations}\textquotedblright. There is no
space to discuss them here, but they can be found in ref.\cite{2tGravity}.

Next, to construct a 2T \textit{field theory} for gravity, a dilaton field
$\Omega\left(  X\right)  $ is also needed in addition to the fields $W\left(
X\right)  ,G_{MN}\left(  X\right)  $ (the field $V_{M}$ can be solved as
$V_{M}=\frac{1}{2}\partial_{M}W,$ so it is not independent). The field theory
action must be such that the \textquotedblleft kinematic
equations\textquotedblright\ mentioned in the previous paragraph must emerge
as some of the equations of motion through the variational
principle\footnote{This is analogous to string theory, where background fields
are restricted by worldsheet local conformal symmetry, while the field theory
must be constructed to reproduce these field equations as equations of motion
derived from the field theory action.}. Furthermore, the Sp$\left(
2,R\right)  $ constraints $Q_{ij}\sim0$ (gauge invariant physical states) must
also be satisfied as \textit{dynamical or kinematical field equations} when
the field interactions are turned off. These requirements, combined with
general coordinate invariance in $d+2$ dimensions, are so strong that they
lead to a unique theory for 2T gravity as a field theory. The action for 2T
gravity is \cite{2tGravity}%
\begin{equation}
S_{grav}=\gamma\int d^{d+2}X\sqrt{G}\left\{
\begin{array}
[c]{c}%
\delta\left(  W\right)  \left[  \Omega^{2}R\left(  G\right)  +\frac{1}%
{2a}\partial\Omega\cdot\partial\Omega-V\left(  \Omega\right)  \right] \\
+\delta^{\prime}\left(  W\right)  \left[  \Omega^{2}\left(  4-\nabla
^{2}W\right)  +\partial W\cdot\partial\Omega^{2}\right]
\end{array}
\right\}  . \label{Sgrav}%
\end{equation}
Here $R\left(  G\right)  $ is the Riemann curvature scalar, $a$ is the special
constant $a\equiv\frac{d-2}{8\left(  d-1\right)  }$, while the potential $V$
can only have the form $V\left(  \Omega\right)  =\lambda\Omega^{\frac{2d}%
{d-2}} $ with a dimensionless coupling $\lambda.$ Other than
$\lambda$ there are no parameters at all. The field $W\left(
X\right)  $ appears in a delta function and its derivative
$\delta\left(  W\right)  $, $\delta^{\prime }\left(  W\right)  ,$ as
well as in additional terms. This unusual and unique structure
emerged from the underlying properties of the Sp$\left(  2,R\right)
$ gauge symmetry on the worldline theory as outlined above. In
particular the delta functions are consistent with one of the
Sp$\left(  2,R\right)  $ physical state requirements $Q_{11}=W\left(
X\right)  =0$, while the others $Q_{12},Q_{22}\sim0$ emerge from the
equations of motion that follow from this action. This field
theoretic structure has a bunch of unusual gauge symmetries of its
own, which I called 2T gauge symmetries \cite{2tGravity}. These are
just strong enough gauge symmetries to eliminate all ghosts from the
2T fields and yield shadows in two lower dimensions (analogs of
Fig.1) that are ghost free physical interacting 1T field theories in
$d$ dimensions. Dualities must relate these shadow 1T field theories
to each other.

This theory of gravity has no dimensionful constants, in particular
there is no Newton's constant $G.$ This emerges from the condensate
of the dilaton (and other scalars, see below) \textit{in the
conformal shadow}. The action above yields a shadow 1T General
Relativity in $d$ dimensions in the form $S_{grav}=\int
d^{d}x\sqrt{-g}\left\{  \phi^{2}R\left(  g\right)  +\frac
{1}{2a}\partial\phi\cdot\partial\phi-V\left(  \phi\right)  \right\}
,$ where $\phi,g_{\mu\nu}$ are the shadows of their counterparts. In
this shadow, due to the special constant $a,$ there is an emergent
local scaling (Weyl) symmetry which is a remnant of general
coordinate transformations in the extra 1+1 dimensions
\cite{2tGravDetails}. Since the coefficient of $R\left( g\right)  $
is positive, the dilaton must have the wrong sign kinetic energy to
satisfy the Weyl symmetry, so $\phi$ is a ghost. Using the Weyl
gauge symmetry the shadow dilaton is gauge fixed to a constant
$\phi_{0}$ (thus eliminating the ghost which would also have been a
Goldstone boson after condensation), yielding precisely Einstein's
General Relativity $S_{grav}=\int d^{d}x\sqrt{-g}\phi_{0}^{2}R\left(
g\right)  $ where the condensate $\phi _{0}^{2}$ must be interpreted
as Newton's constant.

Matter fields can be added, including Klein-Gordon scalars $S_{i}\left(
X\right)  $, Dirac or Weyl spinors $\Psi_{\alpha}\left(  X\right)  $ and
Yang-Mills type vectors $A_{M}\left(  X\right)  $, all in $d+2$ dimensions.
There are special restrictions on each one of these, on the form of their
kinetic energies, and the forms of permitted interactions among themselves and
with the gravitational triplet $\left(  W,\Omega,G_{MN}\right)  .$ These
restrictions emerge from the underlying Sp$\left(  2,R\right)  $ gauge
symmetry and the corresponding physical state conditions at the worldline level.

Within these restrictions I constructed the 2T field theory for the
Standard Model in 4+2 dimensions \cite{2tstandardM}. It is a
perfectly consistent, ghost free 2T field theory because of the new
2T gauge symmetries \cite{2tstandardM} satisfied by these new field
theoretic structures. In the conformal shadow it yields the usual
Standard Model in 3+1 dimensions which is in exquisite agreement
with experiment. This shadow Standard Model has some additional
constraints on the scalar sector (Higgs and others) and their
interaction with the dilaton. The new features are consistent with
known phenomenology, but may help shed some light on Higgs
physics\footnote{This is in spirit similar to the Higgs
phenomenology talk we heard from J. Gunion in this conference (see
also \cite{shapashnikov}\cite{quiros}) because the Higgs in the 2T
Standard Model is required to couple to at least one additional
scalar \cite{2tstandardM}, which may be the dilaton, or another
\textit{electroweak neutral} scalar. Note that historically the
importance of this neutral sector for phenomenlogy was pointed out
based on the prediction from the 2T Standard Model
\cite{2tstandardM} prior to the (less theoretically motivated)
recent phenomenological studies were undertaken.} when more data
becomes available, and perhaps the absence of axions
\cite{2tstandardM} which awaits further clarification until the
\textit{quantum }version of 2T field theory is better understood.

In the coupling of gravity to matter there is another interesting physics
prediction to be emphasized. Every scalar $S_{i}$ in the complete field theory
must couple to the curvature term just like the dilaton in Eq.(\ref{Sgrav}),
but with the opposite sign and standard normalization in the kinetic term.
Then in the conformal shadow the curvature term is predicted to take the form
$(\phi^{2}-a\sum_{i}s_{i}^{2})R\left(  g\right)  $ with a required relative
minus sign! Hence the gravitational constant must emerge from the condensates
of all the scalars, not only the dilaton's. This predicts a physical effect,
that the effective gravitational constant $G\sim(\phi^{2}-a\sum_{i}s_{i}%
^{2})^{-1}$ is not really a constant, rather it must increase after
every phase transition of the universe as a whole (since the
dominant part of each field is the condensate after the phase
transition, this quantity is approximately a constant in between the
phase transitions). Thus the Newton constant we measure today cannot
be the same as the analogous constant before the various transitions
occurred, such as inflation, grand unification, SUSY breaking,
electroweak symmetry breaking. Of course the earlier ones are the
dominant condensates in the sum. There is also the curious
possibility that $G\sim(\phi^{2}-a\sum_{i}s_{i}^{2})^{-1}$ could
turn negative if the other scalars dominate over the dilaton in some
regions of the universe, or in the history of the universe, thus
producing antigravity in those parts of spacetime. In fact, the Big
Bang may be related to the vanishing of $(\phi
^{2}-a\sum_{i}s_{i}^{2}$) at which point the effective $G$ blows up.
The effects of this idea on cosmology is currently under
investigation \cite{2tcosmology}.

\section{Progress in 2T-physics}

Here I will list various points with only very brief comments due to lack of space.

1) Local Sp(2,R) on the worldline, as a gauge symmetry in phase space, has
proven to be a physically correct general principle in both classical and
quantum mechanics. The advantages of this new principle include the
unification of various 1T dynamical systems under a new unifying umbrella
which I called 2T-physics (as summarized in the example of Fig.1). As compared
to 1T-physics, 2T-physics reveals much more \textit{correct information on
physical phenomena} which is systematically missed in the usual formalism of
physics at all scales of distance or energy.

2) The principles of 2T field theory in $d+2$ dimensions have been
established. New types of gauge symmetries eliminate all ghosts and produce a
physical sector which effectively is in two lower dimensions. As in the
corresponding worldline theory, the 2T field theory produces shadow 1T field
theories with duality relations among them (field analogs of Fig.1). Within
these principles I have constructed various physically relevant 2T field
theories. These include the Standard Model, General Relativity, Grand Unified
theories, in 4+2 dimensions, whose conformal shadows are basically the same as
the familiar corresponding theories in $3+1$ dimensions, except for some
additional restriction (mainly on scalar fields) which so far are consistent
with phenomenology, and may even lead to measurable signals at the LHC or in cosmology.

3) Both the worldline and field theory approaches to 2T-physics have been
generalized to supersymmetry. In particular, the general $N=1,2,4$ SUSY 2T
field theories in $4+2$ dimensions have been constructed \cite{susy2tN1}%
\cite{susy2tN2N4}. Consequently, the SUSY generalizations of the
Standard Model or GUTS in $4+2$ dimensions are already available.
So, if SUSY phenomenology becomes relevant at the LHC, the
constraints from 2T-physics
could become interesting.%

\begin{center}
\includegraphics[
natheight=3.474000in,
natwidth=4.624100in,
height=3.474in,
width=4.6241in
]%
{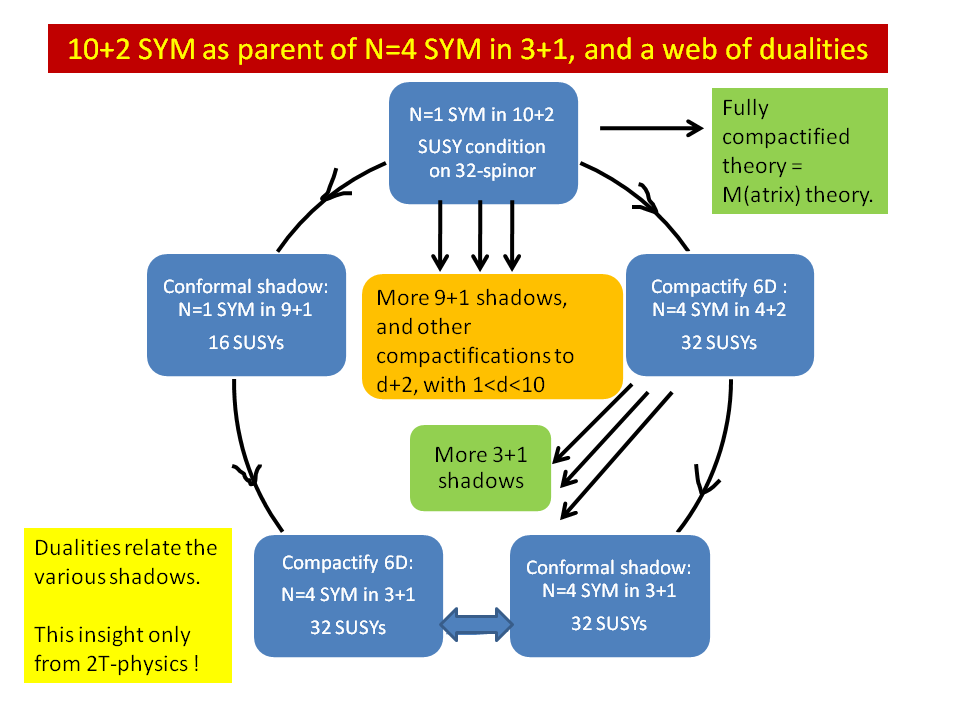}%
\\
Fig.2 - SYM in 12D is the parent of 11D SYM, N=4 SYM in 3+1, N=2 SYM in 4+2,
M(atrix)-theory and others.
\label{12sympng}%
\end{center}

4) SUSY in higher dimensions with 2T has also been achieved. In
particular the super Yang-Mills theory in 10+2 dimensions has been
constructed \cite{12Dsusy}. This 12-dimensional field theory is the
first one to ever go beyond 11 dimensions. It yields many
interesting shadows as well as compactifications that unify various
theories, from M(atrix)-theory to the hotly pursued $N=4$ super
Yang-Mills theory in 3+1 dimensions, and thus may lead to possible
new insights in these 1T theories. These connections are outlined in
Fig.2, and are the basis for considering the new theory as the
parent of all those mentioned in the figure. Finally a connection
between 2T-physics and 12D and 13D S-theory, where it all started in
1995 \cite{Stheory}, is beginning to emerge in a clearer way.

5) Supergravity in 2T-physics has not been constructed yet, but seems to be
around the corner as the path to follow is now clear. The expected maximal
SUSY theory in 13 dimensions should yield 11-dimensional supergravity as the
conformal shadow.

6) For strings, branes, and more generally for M-theory in
2T-physics, there is only old partial progress for tensionless
strings and branes, and more recent progress on the 2T version of
the twistor string \cite{2tstrings}. The usual tensionful string has
historically resisted 2T-physics and the reason for this may have
now become clear: it must be the fact that the tension is
dimensionful, but as I explained above 2T gravity does not allow any
dimensionful constants. The tension will have to emerge from some
condensate. This new insight has not yet been implemented.

7) There is the potential of developing powerful new computational tools that
take advantage of the dualities of the shadows. As in other examples of
dualities, a given theory may be more easily solvable in one dual version as
compared to another. Since the gauge invariant physical content is
holographically captured by every shadow, it may be possible to study physical
effects more easily in some shadows and transform the result to the shadow of
interest (which may be the conformal shadow in the case of field theory). The
shadow phenomena is much more easily handled in the worldline formalism, while
in field theory so far there is limited progress because only some shadows are
easy to study but others seem to be more difficult \cite{emergentfieldth1}%
\cite{emergentfieldth2}. In any case, due to lack of time, very
little has been done so far to take advantage of this feature of
2T-physics, but I think it is where 2T-physics may become most
useful to 1T physicists as well as where most tests of 2T-physics
can be developed.

8) As a final comment, I should mention that I consider all the encouraging
progress in 2T field theory to be only a stepping stone toward a more
comprehensive 2T theory which is based on fields in phase space, not just
position space. I expect the field theory to be non-commutative along the
lines initiated in ref.\cite{noncommutative}. When this approach can be
connected to the successful 2T field theories that work at the present, I
expect much more dramatic insight and progress.

\section{2T-Physics as a unifying framework for 1T-physics}

I would like to conclude this talk with an allegory which may be helpful to
convey the basic idea of 2T-physics and its role as a completion of
1T-physics. I should warn that, the allegory is not perfect and is not a
substitute for the equations. So it should not be taken too far without the
corresponding equations.

As in Fig.3, consider an object in a room. In the allegory this represents
phase space $X^{M},P_{M}$ in $d+2$ dimensions, with the associated Sp$\left(
2,R\right)  $ generators $Q_{ij}\left(  X,P\right)  $. Then consider the
\textit{many} shadows of this object on the surrounding walls which could be
formed by shining light on it from different directions. In the allegory the
shadows represent the \textit{many} emergent physical phase spaces $\left(
x^{\mu},p_{\mu}\right)  $ in two lower dimensions which solve $Q_{ij}=0$, as
in Eqs.(\ref{massless},\ref{massiveRel},\ref{nonrelativistic}) and Fig.1.

An essential point is that one single object in the room (in the allegory a
specific set of $Q_{ij}$, such as Eq.(\ref{special})) has many shadows. To
observers that are stuck on the wall (like we are stuck in 3+1 dimensions) the
various shadows appear like different \textquotedblleft
beasts\textquotedblright\ performing different unrelated motions. However, an
observer in the room immediately knows that the many shadows coming from the
same object must be related. These relationships among shadows can in
principle be discovered by careful observers who live on the wall, but who
have no privilege of being in the room.

The gauge choices in phase space that create the shadows are the analogs of
the many perspectives for observing the object in the room. The relationships
among the emergent dynamical systems in Fig.1 comes from the many forms in
which the same gauge invariant information in $d+2$ dimensions is encoded
\textit{holographically} in each shadow in $d$ dimensions. The 2T-physics
formulation makes the relationships between the \textquotedblleft
shadows\textquotedblright\ evident and predicts them to the 1T physicists on
the \textquotedblleft wall\textquotedblright\ who can study them and verify
them. The information in these relationships, which I called hidden
\textquotedblleft dualities\textquotedblright\ and hidden symmetries, is
information about the perspectives, and therefore it is information that
relates to the properties of the larger space-time in $d+2$ dimensions. By
interpreting this data correctly, and recognizing its relation to the higher
dimensions, the 1T physicists on the \textquotedblleft wall\textquotedblright%
\ have a probe for studying the extra dimensions, albeit indirectly.

In this way 2T-physics provides the privilege of being in the room. It gives
us the ability to recognize that certain systems are indeed related, and that
the predicted relationships are interpreted as perspectives in a higher space-time.%

\begin{center}
\includegraphics[
natheight=2.532200in,
natwidth=2.985300in,
height=2.5322in,
width=2.9853in
]%
{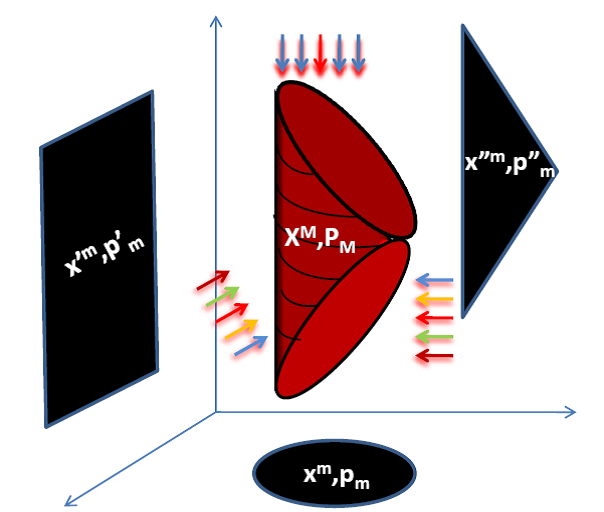}%
\\
Fig.3 - An allegory. \textquotedblleft Room\textquotedblright=4+2,
\textquotedblleft Walls\textquotedblright=3+1. Substance versus shadows.
\label{squareroundtrianglepng}%
\end{center}

Colleagues that have followed my work have generally been in agreement with my
results. Usually I receive encouragement and never criticism. However,
sometimes I am asked: \textquotedblleft All of this is quite nice, but do we
really need 2T, can't we do everything with 1T anyway?\textquotedblright. This
attitude is probably part of the reason for having doubts on whether to invest
effort in 2T-physics.

My answer to this question is a definite, yes you do need 2T, you cannot do
everything with 1T! I have already displayed examples of systems, as in Fig.1,
where new information, not available in 1T-physics is obtained in 2T-physics.
This shows definitely that in principle 1T-physics \textit{systematically}
misses information, while 2T-physics makes it accessible with definite
predictions. 2T-physics opens new avenues and provides new information not
available before. 1T-physics is clearly incomplete. Apparently this has not
been fully appreciated yet by many of my colleagues.

Given that this is a fact, I would guess that, besides seeking practical
applications in 1T-physics, in seeking the unified theory for everything we
may find that the additional information of the extra 1+1 dimensions may lead
to the \textquotedblleft Holy Grail\textquotedblright.

To conclude, I would like to come back to the quotation from Gell-Mann:
\textquotedblleft Anything which is not forbidden is
compulsory!\textquotedblright. Murray used this phrase in connection to the
properties of the strong interactions. But I would like to adopt it to the
fundamental laws of physics.

\end{document}